\documentclass[fleqn, usenatbib]{mnras}

\usepackage{newtxtext,newtxmath}

\usepackage[T1]{fontenc}

\DeclareRobustCommand{\VAN}[3]{#2}
\let\VANthebibliography\thebibliography
\def\thebibliography{\DeclareRobustCommand{\VAN}[3]{##3}\VANthebibliography}

\usepackage{graphicx}	
\usepackage{amsmath}	

\usepackage{amssymb}	

\usepackage{pdflscape}
\usepackage{newtxtext,newtxmath}

\usepackage{mathtools}
\usepackage{mathrsfs}
\usepackage{bm}

\usepackage{ulem}
\usepackage{soul}
\usepackage[dvipsnames]{xcolor}

\title[Engulfment of a hot Jupiter into Kepler-56]{Engulfment of a hot Jupiter as a possible origin of the rapid spin and internal spin misalignment of the planet-hosting red giant Kepler-56}

\author[T.Tokuno]{
Takato Tokuno$^{1,2}$\thanks{E-mail: tokuno-takato@g.ecc.u-tokyo.ac.jp (TT)}
\\
$^{1}$ Department of Astronomy, School of Science, The University of Tokyo, 7-3-1, Hongo, Bunkyo-ku, Tokyo 113-0033, Japan \\
$^{2}$ School of Arts \& Sciences, University of Tokyo, 3-8-1, Komaba, Meguro-ku, 153-8902 Tokyo, Japan \\
}

\date{Accepted XXX. Received YYY; in original form ZZZ}

\pubyear{2025}

\begin{document}
\label{firstpage}
\pagerange{\pageref{firstpage}--\pageref{lastpage}}
\maketitle

\begin{abstract} 
A recent asteroseismic analysis suggests that Kepler-56 – a planet-hosting red giant – exhibits a unique spin structure: (1) the spin axes of the core and envelope are misaligned; and (2) the envelope rotates approximately an order of magnitude faster than typical red giants. In this paper, we investigate a feasible scenario to reproduce this spin structure by estimating the amount of the angular momentum (AM) supply from the planets through the simplified calculation of the time evolution of AM. As a result, unless the tidal efficiency is extremely high, we show that the tidal interactions between the known close-in planets (Kepler-56 b and c) are insufficient to supply the AM required to accelerate Kepler-56 from the spin rate observed in typical red giants. We also show that the engulfment of a hot Jupiter can be expected to provide sufficient AM supply for the acceleration ant that the mass and orbit of the engulfed hot Jupiter are constrained by a mass of 0.5-2 Jupiter masses and an orbital period of 1-6 days. On the other hand, if Kepler 56 was already rapidly spinning before entering the RG stage and requires no acceleration, the obliquity damping by the known close-in planets can reproduce the spin structure of Kepler-56. Even in such cases, planetary engulfment during the MS stage might be involved in achieving rapid spin before the tidal alignment. These discussions demonstrate the importance of Kepler-56 as a candidate for planetary engulfment that may leave traces of its spin structure. 
\end{abstract}




\section{Introduction}

Stellar spin is one of the cornerstones of understanding stellar physics. It plays a crucial role in governing magnetism through the dynamo mechanism and influences chemical mixing in the stellar interior \citep[e.g.][]{Maeder2000ARA&A}. Therefore, the evolution of angular momentum (AM) in a star has been actively discussed, and several theories have been proposed as mechanisms governing the evolution \citep[e.g.][]{Maeder2009Springer}: AM loss due to stellar wind, magnetic instability, meridional circulation, and internal waves.

In addition to these processes, AM exchange with a companion can play a significant role. It is known that orbital AM of the companion can be exchanged with spin AM through tidal or magnetic interactions \citep{Zahn1975A&A, Zahn1977A&A, Hut1981A&A, Benbakoura2019A&A, Ahuir2021A&A}, mass transfer \citep{Packet1981A&A, Dervisoglu2010MNRAS}, merger/engulfment \citep{deMink2013ApJ, Privitera2016aA&A, Privitera2016cA&A, Lau2025A&A}. Generally, these interactions accelerate the stellar spin rate and decrease the inclination between the stellar spin and orbit. In this respect, they are considered to be the origin of observational properties of stellar spin that cannot be explained by the single-star evolution, such as a fast-spinning star \citep{Shao2014ApJ, Maxted2015A&A, TejadaArevalo2021ApJ} or the spin-orbit alignment \citep{Winn2010ApJ, Albrecht2011ApJ, Albrecht2012ApJ, Albrecht2022PASP}. However, the quantitative nature of these processes remains poorly constrained.

For understanding stellar AM evolution, red giant (RG) stars provide important insights because asteroseismic analysis yields great constraints on their spin profile \citep[e.g.][]{Aerts2019ARA&A}. The radiative core and convective envelope in RG behave as $g$-mode and $p$-mode cavities, respectively. The rotational splitting of gravito-acoustic mixed mode enables estimation of the spin periods of both the core and envelope  \citep{Beck2012Natur, Mosser2012A&A, Deheuvels2014A&A}. Moreover, the amplitudes of the split modes give a constraint on the inclination angle of the spin axis. In fact, the long-baseline and high-precision photometry obtained by the \textit{Kepler} mission \citep{Borucki2010Sci} provided spin profiles for well over a thousand red giants \citep{Gehan2018A&A, Li2024A&A, Dhanpal2025arXiv}, enabling more detailed verification of the stellar AM evolution \citep{Fuller2019MNRAS, Eggenberger2019A&A,Takahashi2021A&A}.

Among them, this paper focuses on the RG star Kepler-56 (KIC 6448890, KOI-1241), which hosts multiple exoplanets and exhibits an unusual spin profile. The transit and radial‐velocity methods revealed that Kepler-56 possesses two close‐in planets, Kepler-56 b and Kepler-56 c \citep{Steffen2013MNRAS, Huber2013Sci, Weiss2024ApJS}. In addition, asteroseismic analysis revealed that the inclination angle of the spin axis relative to the line of sight (hereafter, spin inclination angle) is about 40 degrees in Kepler-56 \citep{Huber2013Sci}. As the transiting planetary orbits are almost edge‐on, this spin inclination implies the spin–orbit misalignment of the Kepler-56 system, which attracts attention to Kepler-56 \citep{Li2014ApJ, Fellay2021A&A}. 

In recent years, the uniqueness of Kepler-56 has been further highlighted by the up-to-date analysis of \citet{Ong2025ApJ}. They constructed an asteroseismic prescription for cases where the spin inclination angle of the core and envelope are independent, while the previous analyses have assumed that they are aligned. Their reanalysis of oscillation data for Kepler-56 found Kepler-56's spin profile different from the previous results \citep[][]{Huber2013Sci, Fellay2021A&A}. Most notably, the spin inclination angle of the envelope is markedly misaligned with that of the core, which is the first evidence of internal spin misalignment. Furthermore, it is suggested that Kepler-56 has a spin rate several times faster than other RGs with similar mass. 

These characteristics of Kepler-56's spin, internal spin misalignment and the rapid envelope spin rate, are difficult to reproduce under the single-star evolution \citep[see also the discussions of][]{Ong2025ApJ}. Thus, it is considered that Kepler-56 is a suitable object for investigating the impact of AM transfer with its companion on the stellar spin. However, the discussion regarding the formation process is still in progress. 

The purpose of this paper is to explore the formation scenario of Kepler-56's spin profile, assuming it to be a by-product of interaction with the planet. Through the observational  constraints and estimation of the AM supply, we investigate a formation scenario for Kepler-56's unusual spin profile. Our discussion is expected to contribute to elucidating the formation process of not only Kepler-56 but also other RGs with the rapid spin or internal spin misalignment, both known and yet to be discovered. 

The rest of the paper is organised as follows: First, we introduce our scenarios and method in Section \ref{sec:method}; Next, we provide the feasibility of the scenarios in Section \ref{sec:result}. Finally, we provide the discussion and summary in Sections \ref{sec:discussion} and \ref{sec:conclusion}.

\section{Method} \label{sec:method}
In this section, we first present a brief summary of the properties of the Kepler-56 system (Section \ref{subsec:meth-properties}). Next, we introduce two scenarios to reproduce the spin profiles of Kepler-56 (Section \ref{subsec:meth-scenario}). Finally, we explain the process to investigate the feasibility of two scenarios (Section \ref{subsec:formulation}). 

\subsection{Properties of the Kepler-56 system}
\label{subsec:meth-properties}

\begin{table}
    \centering
    \caption{Parameters of the Kepler-56 system adopted in this paper.}
    \label{tab:values_K56}
    \setlength\tabcolsep{0.1cm}
    \begin{tabular}{lccc} 
    \hline
    \textit{Kepler-56 (Star)} \\
    \hline
    Parameters &  Symbol  & Adopted Value$^{*a}$ & Source$^{*b}$  \\
    \hline
    Stellar mass & $M_\mathrm{s}$ & $1.32 \pm 0.13 \, \mathrm{M}_\odot$ & (1) \\
    Stellar radius & $R_\mathrm{s}$ & $4.23 \pm 0.15 \, \mathrm{R}_\odot$ & (1) \\
    Effective temperature & $T_\mathrm{eff}$ & $4840 \pm 97 \, \mathrm{K}$ & (1) \\
    Surface gravity (cgs) & $\log g$ & $3.31 \pm 0.01$ & (1) \\
    Metallicity & [Fe/H] & $0.20 \pm 0.16$ dex & (1) \\
    Moment of inertia & $I_\mathrm{s}$ & $3.43^{+0.72}_{-0.64} \, \mathrm{M}_\odot \mathrm{R}_\odot^2$ & (2) \\
    Spin period (env.) & $P_\mathrm{env}$ & $74 \pm 3 \, \mathrm{d}$ $^{*c}$ & (1)(3) \\
    Spin inclination angle (env.) & $i_\mathrm{env}$ & $104 \pm 4 \, \mathrm{deg}$ & (3) \\
    Spin period (core) & $P_\mathrm{core}$ & $21.3 \pm 0.2  \, \mathrm{d}$ & (3) \\
    Spin inclination angle (core) & $i_\mathrm{core}$ & $43 \pm 2  \, \mathrm{deg}$ & (3) \\
    \hline
    \hline
    \textit{Kepler-56 b (Planet)} \\
    \hline
    Parameters &  Symbol  & Value & Source  \\
    \hline
    Planetary mass & $M_\mathrm{pl, b}$ & $0.11 \pm 0.02 \, \mathrm{M}_\mathrm{J}$ & (1)(4) \\
    Orbital period & $P_\mathrm{orb, b}$ & $10.50 \, \mathrm{d}$ & (1) \\
    \hline
    \hline
    \textit{Kepler-56 c (Planet)} \\
    \hline
    Parameters & Symbol & Value & Source \\
    \hline
    Planetary mass & $M_\mathrm{pl, c}$ & $0.74 \pm 0.03 \, \mathrm{M}_\mathrm{J}$ & (1)(4) \\
    Orbital period & $P_\mathrm{orb, c}$ & $21.40 \, \mathrm{d}$ & (1) \\
    \hline
    \multicolumn{4}{l}{$^{*a}$ $\mathrm{M}_\odot$ and $\mathrm{M}_\mathrm{J}$ denote the solar mass, and Jupiter mass, respectively.} \\
    \multicolumn{4}{l}{Also, $\mathrm{R}_\odot$ denotes the solar radius.} \\
    \multicolumn{4}{l}{$^{*b}$ Reference list: (1) \citet{Huber2013Sci}; (2) This work (Section \ref{subsec:meth-Lacc}); } \\ 
    \multicolumn{4}{l}{(3) \citet{Ong2025ApJ}; (4) \citet{Weiss2024ApJS} } \\
    \multicolumn{4}{l}{$^{*c}$ The value is from the results of spot modulation \citep{Huber2013Sci}.} \\
    \end{tabular} 
\end{table}

\begin{figure}
\centering
 \includegraphics[width=\columnwidth]{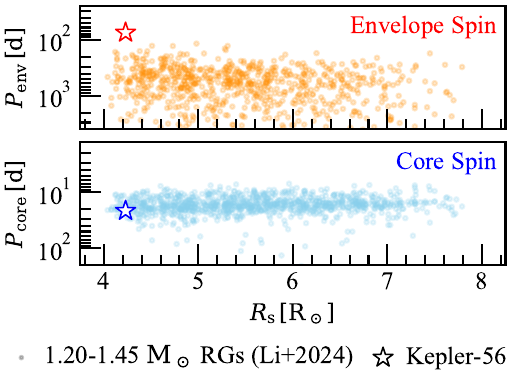}
 \caption{Comparison of the spin rates in envelope (ordinate of top panel) and core (ordinate of bottom panel) between Kepler-56 and other RGs \citep{Li2024A&A} with 1.20–1.45 $\mathrm{M}_\odot$. The abscissae in both panels are the stellar radius. Kepler-56 is marked in the star-shaped symbol, while other RGs are marked in the circle symbols. }
 \label{fig:Prot_env_core}
\end{figure}

In Table \ref{tab:values_K56}, we summarise the symbols used in this paper and the observed values in the Kepler-56 system. This table and the discussion below are based on \citet{Huber2013Sci}, \citet{Weiss2024ApJS} and  \citet{Ong2025ApJ}. 

\subsubsection{Stellar properties}
\label{subsec:properties-star}

First, we introduce the stellar properties of Kepler-56. The joint analysis of asteroseismic and spectroscopic observations indicates that Kepler-56 is the star at the base of the RG branch ($T_\mathrm{eff} = 4840 \pm 97 \,\mathrm{K}$ and $\log g = 3.31 \pm 0.01$). Also, the estimated mass and radius are $1.32 \pm 0.13 \, \mathrm{M}_\odot$ and $4.23 \pm 0.15 \, \mathrm{R}_\odot$, respectively, indicating that Kepler-56 is an RGs originating from an F-type main-sequence (MS) star. Also, it is revealed that Kepler-56 is metal-rich, with [Fe/H] $=+0.20 \pm 0.16$.

The most remarkable feature of Kepler-56 is its unique spin profile. An asteroseismic analysis by \citet{Ong2025ApJ} uncovered the structured spin profile of Kepler-56: the core spin with a period of $P_\mathrm{core} = 21.3 \pm 0.2 \, \mathrm{d}$ and an inclination angle of $i_\mathrm{core} = 43 \pm 2 \, \mathrm{deg}$; the envelope with a period of $P_\mathrm{env} = 72^{+5}_{-8}$ days and $i_\mathrm{env} = 104 \pm 4 \, \mathrm{deg}$. These spin inclination angles indicate that the true obliquity of the spin axis between the core and envelope is at least 60 degrees or more, considering the effect of projection in the direction of sight. 

Furthermore, compared to RGs of similar mass, the envelope spin of Kepler-56 is  noticeably rapid. Fig.~\ref{fig:Prot_env_core} shows the comparison of $P_\mathrm{env}$ and $P_\mathrm{core}$ between Kepler-56 and other RGs with similar masses, which are 1.20–1.45 $\mathrm{M}_\odot$ \citep{Li2024A&A}. It is found that,  while $P_\mathrm{core}$ of Kepler-56 remains within the typical range observed for RGs, $P_\mathrm{env}$ of Kepler-56 is several times more rapid than those of other RGs. 

The surface spin rate of Kepler-56 measured by other methods is consistent with the asteroseismology-based spin period in the envelope, $P_\mathrm{env} \approx 70 \mathrm{d}$. One is based on the Lomb–Scargle analysis for the spot-induced quasi-periodic variations in the light curves, suggesting a surface spin rate of $74 \pm 3 \, \mathrm{d}$ \citep{Huber2013Sci}\footnote{Since the spin rate from the spot modulation is more accurate, this paper adopts it as the envelope spin rate (see Table \ref{tab:values_K56})}. The other is measurements of spectral broadening from spectroscopic analysis, which support $ v \sin i = 2$–$3 \, \mathrm{km \cdot s^{-1}}$ \citep{Petigura2017AJ, Rainer2023A&A}, where $v \sin i$ is a rotational velocity. This is in agreement with the estimated value from asteroseismic results, $v \sin i = 2.8 \pm 0.3 \, \mathrm{km \cdot s^{-1}}$. 

Since the two methods above and the asteroseismic method present the spin rate at the surface and the kernel-weighted envelope spin rate, respectively, the consistency shown above supports uniform spin within the envelope \citep{Ong2025ApJ}. It enables us to calculate the total AM of Kepler-56 (Section \ref{subsec:meth-Lacc}).

\subsubsection{Planetary properties}
\label{subsec:properties-planet}

Second, we introduce the properties of planets around Kepler-56. Orbiting Kepler-56 are two close-in gaseous planets, Kepler-56 b and Kepler-56 c, both of which transit the stellar disc. The radial velocity observations \footnote{To be precise, \citet{Weiss2024ApJS} indicate the minimum mass, but considering the orbital inclination suggested by the transit \citep{Huber2013Sci}, we use it as the mass. } reveal that an inner planet, Kepler-56 b, is a  Neptune-sized planet ($M_\mathrm{pl, b} = 0.11 \pm 0.04 \, \mathrm{M}_\mathrm{J}$) with an orbital period of $P_\mathrm{orb,b} = 10.5 \, \mathrm{d}$, while an outer planet, Kepler-56 c, is a sub-Jovian planet ($M_\mathrm{pl, c} = 0.74 \pm 0.03 \,\mathrm{M}_\mathrm{J}$) with $P_\mathrm{orb,c} = 21.4 \, \mathrm{d}$. 

Both planets reside in nearly circular and coplanar orbits. Since they have transiting orbits, their orbital inclination angles are perpendicular to the line of sight. It suggests that, at least in terms of projected inclination, their orbital axes tend to align with the spin axis in the envelope. Conversely, for core spin, there is spin-orbit misalignment by at least 50 deg.

We note that the radial‐velocity observations show the presence of a distant outer planet, Kepler-56 d, and another planet further outside \citep{Otor2016AJ,Weiss2024ApJS}. It is suggested that the torque caused by these distant planets can affect the orbital planes of Kepler-56 b and c \citep{Huber2013Sci, Li2014ApJ}. In our calculations, which focus only on the post–main-sequence (post-MS) stage, this effect is considered negligible. However, its possible impact is discussed in Section \ref{subsec:rapid-spin}, which discusses the long-term evolution including the MS stage.

\subsection{Our formation scenarios}
\label{subsec:meth-scenario}

\begin{figure*}
 \includegraphics[width=2\columnwidth]{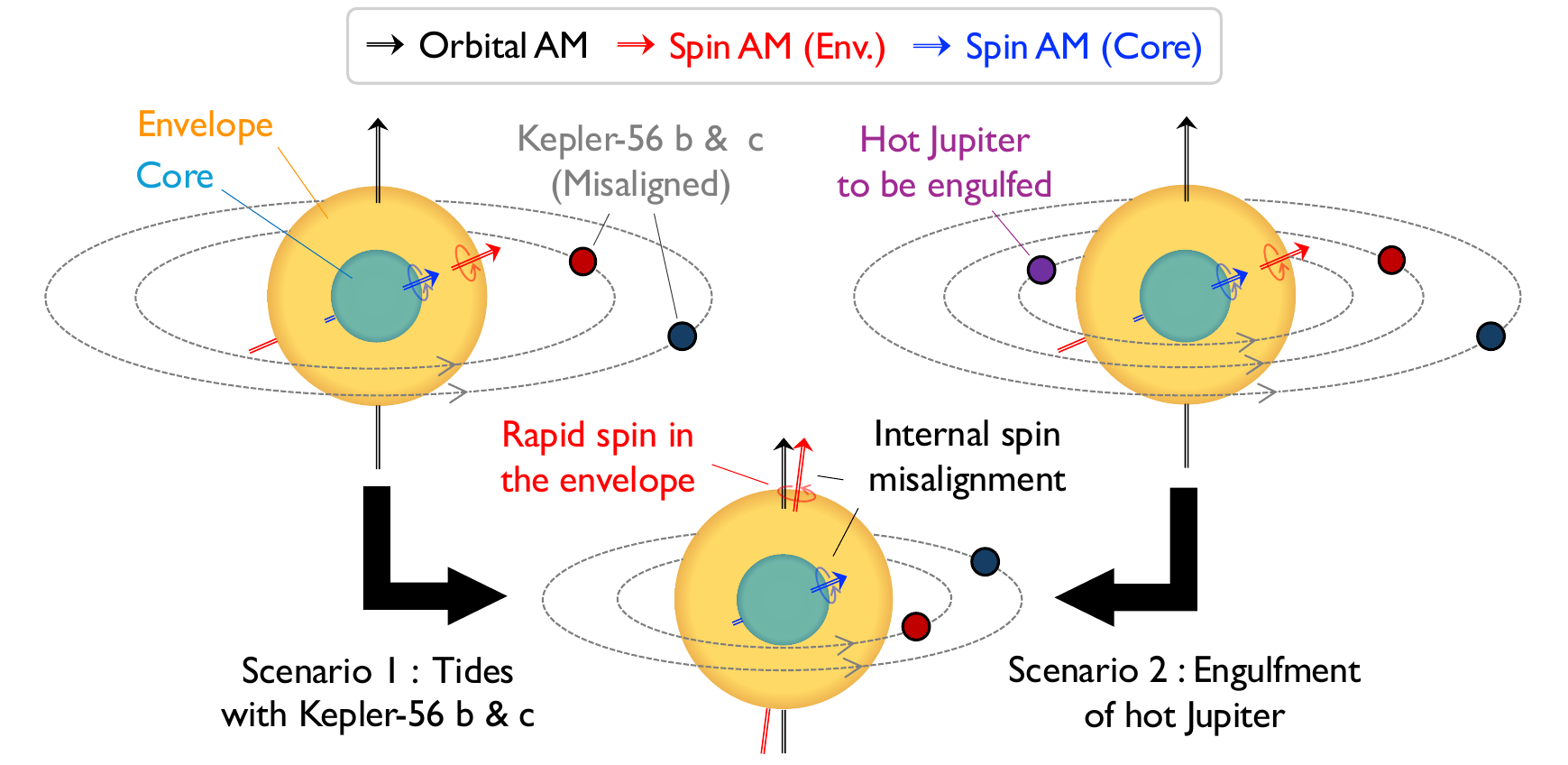}
 \caption{Schematic picture of our scenario to reproduce the spin profile of Kepler-56. Initially, the spin axes of the core and envelope were aligned in a direction offset from the orbit. After AM supply from the planetary orbit to the stellar envelope, the envelope spins more rapidly in a direction offset from the core. This AM supply is considered to occur during the post-MS stage (refer to the main text for details).}
 \label{fig:schematic}
\end{figure*}

We consider the initial condition that the original spin axes of the core and the envelope are aligned but misaligned with the planetary orbit. When the stellar spin is faster than the planetary revolution, the tidal interactions transports the planet's orbital AM to the stellar spin. This causes the stellar spin to accelerate, the orbital radius to decrease, and the spin-orbit obliquity to decrease, simultaneously. Given that this effect works on the envelope, this process may reproduce the spin structure of Kepler-56. To clarify, in this study we consider tidal processes working in the convective envelope, namely the equilibrium tide \citep{Zahn1989A&A, Mathis2016A&A} and the dissipation of the inertial waves \citep{OgilvieLin2007ApJ,Lai2012MNRAS,Ogilvie2013MNRAS} or $f$-modes \citep{Vick2020MNRAS}. By contrast, we do not consider the dissipation of $g$-modes \citep{BarkerOgilvie2010MNRAS, Essick2016ApJ, MaFuller2021ApJ, Weinberg2024ApJ}, as it is expected primarily to transfer AM to the radiative core. This is supported by the fact that the spin rate of Kepler-56's core is comparable to other RGs despite the its small moment of inertia (Section \ref{subsec:properties-star}).

On the basis of this, we consider that two scenarios can explain the current state of the Kepler-56 system (see Fig.~\ref{fig:schematic} for a schematic picture). The first one is that the required AM is supplied by tidal interactions with the known planets Kepler-56 b and c \citep[][]{Winn2010ApJ,Meynet2017A&A, Saunders2024AJ}. This scenario can reproduce the current property where the envelope is more closely aligned with the orbits of Kepler-56 b and c than with the core. 

The second one is the AM supply through the engulfment of a hot Jupiter (HJ) that originally existed \citep[][see Section \ref{subsec:companion} for the possibility of heavier companions]{Carlberg2013AN, Tayar2022ApJ, O'Connor2023ApJ, Lau2025A&A}. Assuming that the orbit of engulfed HJ is coplanar with Kepler-56 b and c but misaligned with the original spin-axis, AM supply through the engulfment can reproduce the spin structure of Kepler-56. Also, it is considered that the engulfed HJ was originally located inside Kepler-56 b in terms of dynamical stability.  

In either scenario, the AM supply is considered to occur during the post-MS stage rather than the MS stage. It is supported by the following considerations \citep[see also the discussion of][]{Ong2025ApJ}. It is generally known that the difference in spin rates between the core and envelope is small during the MS star \citep{Kurtz2014MNRAS, Saio2015MNRAS, Benomar2015MNRAS, VanReeth2018A&A, Saio2021MNRAS}. Then, if there is massive AM supply to the envelope, the envelope rotates faster than the core, causing realignment of the spin axis due to internal AM transport from the envelope to the core. On the other hand, in the post-MS stage, the difference in spin rates between the core and envelope is large, and the moment of inertia of the envelope is also large. In such cases, the envelope spins slower than the core even after the massive AM supply, as in Kepler-56. This profile makes it easier to maintain internal spin misalignment because the direction of internal AM transport is from the core to the envelope. 

Furthermore, the picture of AM supply occurring in the post-MS stage is consistent with the trend of the efficiency of tidal torque. The tidal torque is proportional to the fifth power of the stellar radius, and therefore becomes extremely strong in the post-MS stage (see Section \ref{subsec:formulation}). Also, compared to F-type MS star with very thin or no surface convective layers, post-MS stars with thick surface convective layers exhibit strong tidal torques \citep{Winn2010ApJ, Lai2012MNRAS, Rogers2013ApJ, Saunders2024AJ}. 

\subsection{Process for investigating feasibility}
\label{subsec:formulation}

We introduce the process for investigating the feasibility of the two scenarios discussed in Section \ref{subsec:meth-scenario} from the perspective of whether AM required to reproduce Kepler-56 can be supplied.

\subsubsection{Estimation of needed AM supply}
\label{subsec:meth-Lacc}

\begin{figure}
 \includegraphics[width=\columnwidth]{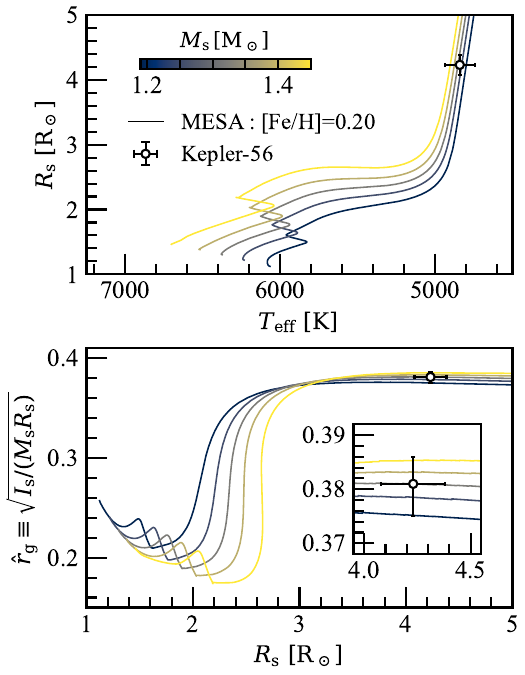}
 \caption{Comparison of Kepler-56 (points with error bars) and MESA evolutionary tracks for the stars with $M_\mathrm{s} = 1.20 -1.45 \, \mathrm{M}_\odot$ and $\mathrm{[Fe/H]} = 0.20$ (coloured lines). The colours of the lines correspond to the stellar mass shown in the colour bar inside the top panel. (Top) Comparison of them in the HR diagram. (Bottom) Gyration radius, $\hat{r}_g \equiv \sqrt{I_\mathrm{s}/(M_\mathrm{s}R_\mathrm{s})}$,  as a function of stellar radius. The plotted gyration radius of Kepler-56 ($\hat{r}_g=0.381^{+0.005}_{-0.006}$) is estimated from its radius. The inset shows a close-up of the position of Kepler-56.}
 \label{fig:MESA-calc}
\end{figure}

We discuss how much AM is needed to reproduce Kepler-56 under the scenario shown in Fig.~\ref{fig:schematic}. 

Here, we denote the spin AM vector observed in Kepler-56 as $\bm{J}_\mathrm{obs}$ and the AM vector in the initial condition as $\bm{J}_\mathrm{init}$. In this case, the AM vector required to reproduce Kepler-56 is $\bm{J}_\mathrm{obs} - \bm{J}_\mathrm{init}$. This AM vector can be separated into the expansion (i.e., spin acceleration) and rotation (i.e., obliquity damping) of the spin AM vector, which are represented as  $\bm{L}_\mathrm{acc}$ and $\bm{L}_\mathrm{damp}$ hereafter. The norms of $\bm{L}_\mathrm{acc}$ can be written as 
\begin{gather}
    L_\mathrm{acc} = J_\mathrm{obs} - J_\mathrm{init} = \alpha J_\mathrm{obs} \label{eq:def-Lacc}, 
\end{gather}
where we introduce the notation $x=|\bm{x}|$ for any vector $\bm{x}$. Also, $\alpha$ introduced here is the coefficient when $L_\mathrm{acc}$ is normalised by $J_\mathrm{obs}$. On the other hand, the norm of $\bm{L}_\mathrm{damp}$ depends on how obliquity damping works and satisfies $2 \sin(\psi/2) J_\mathrm{init} \leq L_\mathrm{damp} \leq 2 \sin(\psi/2) J_\mathrm{obs}$, where $\psi$ is the angle formed by $\bm{J}_\mathrm{obs}$ and $\bm{J}_\mathrm{init}$. 

The specific value of $L_\mathrm{acc}$ is estimated as follows. We estimated the value of $J_\mathrm{obs}$ using the stellar evolution code MESA \citep{Paxton2011ApJS}. Using a stellar evolution tracks that reproduce Kepler-56 on the HR diagram (our setting is shown in Appendix \ref{app:MESA}), we determined the gyration radius, $\hat{r}_g \equiv \sqrt{I_\mathrm{s}/(M_\mathrm{s}R_\mathrm{s})}$, where $I_\mathrm{s}$ is the stellar moment of inertia. Fig.~\ref{fig:MESA-calc} displays the HR diagram (top panel) and relationship  between $R_\mathrm{s}$ and $\hat{r}_g$ (bottom panel) of our evolutionary tracks, showing $\hat{r}_g=0.381^{+0.005}_{-0.006}$ at the current status of Kepler-56. Thus, we can estimate $I_\mathrm{s} = 3.43^{+0.72}_{-0.64} \, \mathrm{M}_\odot \mathrm{R}_\odot^2$ from Kepler-56's $M_\mathrm{s}$ and $R_\mathrm{s}$ (Table \ref{tab:values_K56}) \footnote{The error range is overestimated by treating the mass and radius uncertainties as independent. Although directly fitting the observed values could reduce the error by several tens of percent, we did not pursue this owing to the limited precision of one significant digit and remaining uncertainties (Section \ref{subsec:addtional-effect}).}. Assuming a uniform envelope of $P_\mathrm{env} = 74 \pm 3 \, \mathrm{d}$ (Section \ref{subsec:properties-star}), $J_\mathrm{obs}$ is estimated to be $2.7$-$3.9 \times 10^{49} \, \mathrm{erg \cdot s}$. Here, we consider the envelope spin AM as the total spin AM for current Kepler-56 since the core spin AM is negligible. On the other hand, Fig.~\ref{fig:Prot_env_core} shows RGs with a radius comparable to Kepler-56 satisfy $P_\mathrm{env} \lesssim 150 \, \mathrm{d}$, inferring that $\alpha$ is in the range of 0.5–1.0. Thus, we obtain $L_\mathrm{acc}=1$-$4 \times 10^{49} \, \mathrm{erg \cdot s}$ from Equation (\ref{eq:def-Lacc}). 

In the scenario reproducing Kepler-56 shown in Section \ref{subsec:meth-scenario}, the spin acceleration corresponding to $L_\mathrm{acc}$ is required. This is the criterion for the feasibility discussion hereafter (Sections \ref{subsec:meth-case1} and \ref{subsec:meth-case2}). The case where acceleration is not required is discussed in Section \ref{subsec:rapid-spin}.

\subsubsection{AM supply through tides with Kepler-56 b and c}
\label{subsec:meth-case1}

In this section, we explain the process to estimate the amount of AM supply used for spin acceleration through the tidal interaction between Kepler-56 b and c, as presented in the first scenario in Section \ref{subsec:meth-scenario}. 

Under the general prescription, the time evolution of AM in the star-planet system through the tidal interaction is described by 
\begin{gather}
    \frac{\mathrm{d} \bm{h}_i }{\mathrm{d}t} = - \mathbf{\Gamma}_{\mathrm{tide},i} , \label{eq:hi-evo-vector} 
\end{gather}
under the conservation of total AM as
\begin{gather}
    \bm{J} + \sum_i \bm{h}_i  = \text{Const.} \label{eq:J-evo-vector}
\end{gather}
Here, $t$ is the time, $\bm{J}$ is the spin AM vector of the host star, $\bm{h}_i$ is the orbital AM vector of the $i$-th planet and $\mathbf{\Gamma}_{\mathrm{tide},i}$ is the torque vector through the tidal interaction between the host star and $i$-th planet. We note that the spin AM vector of the planet is neglected in this formulation because the planetary spin is easily synchronised \citep{Guillot1996ApJ}. Also, we neglect the AM loss due to magnetic braking (the validity is discussed in Section \ref{subsubsec:mag-braking}). 

Assuming that all planetary orbits are circular and coplanar, which are satisfied in the observed Kepler-56 system (see Section \ref{subsec:meth-properties}), the norm of the orbital AM vector is described as 
\begin{gather}
   |\bm{h}_i| =  \frac{M_\mathrm{s}M_{\mathrm{pl},i} G^{2/3}}{(M_\mathrm{s} + M_{\mathrm{pl},i})^{1/3}} n_i^{-1/3}, \label{eq:def-hi-norm}
\end{gather}
where $n_i = 2\pi/P_\mathrm{orb,i}$ denotes the angular velocity of the $i$-th planetary orbit, and $G$ is the gravitational constant. Also, under this assumption, the norm of tidal torque can be formulated by using the dimensionless quality factor of the star, $Q'$, as 
\begin{gather}
    |\mathbf{\Gamma}_{\mathrm{tide},i}| =  \frac{R_\mathrm{s}^5 n_i^4}{G} \left( 1 + \frac{M_\mathrm{s}}{M_{\mathrm{pl},i}} \right)^{-2} \left( \frac{9}{4 Q'} \right), \label{eq:def-Gamma_tide-norm}
\end{gather}
in a traditional framework \citep{Goldreich1966Icar, MurrayDelmott1999}. We note that the physical definition of $Q'$ is inversely proportional to the phase lag between the tidal potential and the tidal bulge divided by the Love number \citep[see e.g.][]{Ogilvie2014ARAA}.

However, there is an uncertainty regarding the expression of $Q'$ and the direction in which $\mathbf{\Gamma}_{\mathrm{tide},i}$ points with respect to $\bm{J}$ and $\bm{h}_i$, both of which are needed to solve Equations (\ref{eq:hi-evo-vector}) and (\ref{eq:J-evo-vector}).
The expression of the $Q'$ is currently the subject of debate in terms of theory \citep{Ogilvie2014ARAA, BolmontMathis2016CeMDA, Barker2020MNRAS, Barker2026enap} and observation \citep{MiyzakiMasuda2023AJ, Adams2024PSJ, Tokuno2024ApJ, Millholland2025ApJ}. Also, while there are various suggestions about the direction of the $\mathbf{\Gamma}_{\mathrm{tide},i}$ vector \citep{Hut1981A&A, Barker2009MNRAS, Lai2012MNRAS, Rogers2013ApJ, Anderson2021ApJ}, it is fair to say that there is no consensus. 

On the basis of the above uncertainties, this study applies three simplifications to focus on the AM supply used for spin acceleration, which is comparable to $L_\mathrm{acc}$ (Section \ref{eq:def-Lacc}). First, we assume that the value of $Q'$ is constant, which is a classical expression \citep{Goldreich1966Icar, MurrayDelmott1999}. This constant $Q'$ model can provide comparable results to non-constant $Q'$ models \citep{Penev2014PASP, Penev2018AJ}. Second, we assume that spin-orbit alignment occurs immediately, which means $\bm{J}$, $\bm{h}_i$ and $ \mathbf{\Gamma}_{\mathrm{tide},i}$ are all parallel. This assumption is supported by observational and theoretical evidence that the efficiency of obliquity damping is much higher than that of orbital decay \citep{Lai2012MNRAS, Saunders2024AJ}. Third, we assume that the tidal torque $\mathbf{\Gamma}_{\mathrm{tide},i}$ always works in the direction of reducing $\bm{h}_i$ ,which is equivalent to assuming that the stellar spin is always slower than the planetary revolution. Since this is not always the case, this approximation gives the maximum value of AM supply. Considering the assumption that $\bm{J}$ and $\bm{h}_i$ are parallel, $\mathbf{\Gamma}_{\mathrm{tide},i}$ is used entirely for spin acceleration (Equation \ref{eq:J-evo-vector}).

By integrating Equations (\ref{eq:hi-evo-vector}), (\ref{eq:def-hi-norm}) and (\ref{eq:def-Gamma_tide-norm}) under these assumptions, we can calculate the maximum amount of AM supply used for spin acceleration, $L_\mathrm{calc, max}$, which is defined as
\begin{gather}
    L_\mathrm{calc, max} = \int^0_{-T} \sum_i |\mathbf{\Gamma}_{\mathrm{orb},i}| \mathrm{d}t, \label{eq:def-Lcalcmax}
\end{gather}
where we set the post-MS stage to $t=[-T,0]$. 
The strength of these approximations is that they reduce the time-dependent stellar parameter to only $R_\mathrm{s}$. In other words, $L_\mathrm{calc, max}$ is determined solely by $R_\mathrm{s}$ and planetary parameters. Since the problem reduces to whether this falls within the range of $L_\mathrm{acc}$ (Section \ref{subsec:meth-Lacc}), there is no need to solve the  stellar AM evolution in post-MS stage, which includes the uncertainties of internal AM profile \citep[e.g.][]{Aerts2019ARA&A}.

Using the parameters of Kepler-56 b and c  (Table \ref{tab:values_K56}) and the initial positions that reproduce the current status, we calculate $L_\mathrm{calc, max}$ for each $Q'$. Competing $L_\mathrm{calc, max}$ and $L_\mathrm{acc}$ enables us to estimate the maximum value of $Q'$, i.e. the minimum tidal efficiency, necessary to reproduce the spin profile of Kepler-56 through the tidal interaction with Kepler-56 b and c. We can examine the feasibility of this scenario by comparing the results obtained with the range of $Q' \simeq 10^4$-$10^8$, which is suggested by previous observations  \citep{Penev2018AJ, Patra2020AJ, Vissapragada2022ApJ, Wong2022AJ, MiyzakiMasuda2023AJ, Adams2024PSJ, Tokuno2024ApJ, Millholland2025ApJ}\footnote{Recent studies have questioned the observation of Kepler-1658, the star corresponding to the upper limit of this range \citep{Winn2025arXiv}. Taking this into account, the range becomes $Q' \simeq 10^5$-$10^8$.}. These observations are of MS stars and subgiants; nevertheless, because Kepler-56 is in an early RG stage and shares key properties, such as comparable mass and the presence of a convective envelope, they serve as a useful basis for comparison.

Regarding the stellar parameters required for calculations, we adopt the evolution tracks for $1.32 \, \mathrm{M}_\odot$ stars calculated by \textsc{mesa} (see Appendix \ref{app:MESA})\footnote{The same calculation was performed using the track of $1.20$-$1.45 \, \mathrm{M}_\odot$, but the results did not change significantly.}. The time interval is defined as the period between the start of expansion, which is after the contraction at the end of the MS stage (`hook'), and the moment when the $R_\mathrm{s}$ reaches that of Kepler-56 (Table \ref{tab:values_K56}). Also, we assume that all planetary masses are constant (we discuss the validity in Section \ref{subsubsec:escape}).

\subsubsection{AM supply through engulfment of hot Jupiter}
\label{subsec:meth-case2}


In this section, we discuss AM supply in the second scenario in Section \ref{subsec:meth-scenario}, where an HJ that originally existed is engulfed. 

As discussed in Section \ref{subsec:meth-case1}, we integrate Equations (\ref{eq:hi-evo-vector}), (\ref{eq:def-hi-norm}) and (\ref{eq:def-Gamma_tide-norm}) for Kepler-56 b and c, and the engulfed planet. We calculate the amount of AM supplied by the engulfment assuming that the orbital AM is entirely transported to the spin AM when the semi-major axis becomes equal to the stellar radius \citep[][]{Lau2025A&A}. 

As parameters, the mass $M_\mathrm{pl,eng}$ and the orbital period $P_\mathrm{orb}$ at the initial conditions for the engulfed HJ are added. For each $Q'$, we can determine the range of $(M_\mathrm{pl,eng}, P_\mathrm{orb,eng})$ where $L_\mathrm{acc}$ can be supplied and the added planet is engulfed during the time interval. Comparing this range of $(M_\mathrm{pl,eng}, P_\mathrm{orb,eng})$ with the observed population of HJs around F-type MS stars, we can discuss the feasibility of this scenario. 

The observed population of HJs is retrieved from the NASA Exoplanet Archive \citep{Akeson2013PASP, Christiansen2025arXiv}. From this catalogue of observed planetary systems, we extract the planets satisfying the following conditions: $M_\mathrm{s} = 1.20$-$1.45 \, \mathrm{M}_\odot$, $M_\mathrm{pl} \geq 0.1 \, \mathrm{M_J}$, $P_\mathrm{orb} \leq 10.5 \, \mathrm{d}$, and the host star is in the MS stage. The third condition corresponds to the requirement that the engulfed HJ should orbit inside Kepler-56 b (Section \ref{subsec:meth-scenario}). In particular, since we present scenarios in which the planetary orbits are misaligned with the stellar spin (Section \ref{subsec:meth-scenario}), we extracted HJs whose projected obliquity measured by the Rossiter–McLaughlin effect exceeds 50 degrees (see also Appendix \ref{app:Obliquity}). 

\section{Result} \label{sec:result}

\subsection{Reproducibility of Kepler-56 by tides in Kepler-56 b and c} \label{subsec:res-case1}

\begin{figure}
 \includegraphics[width=\columnwidth]{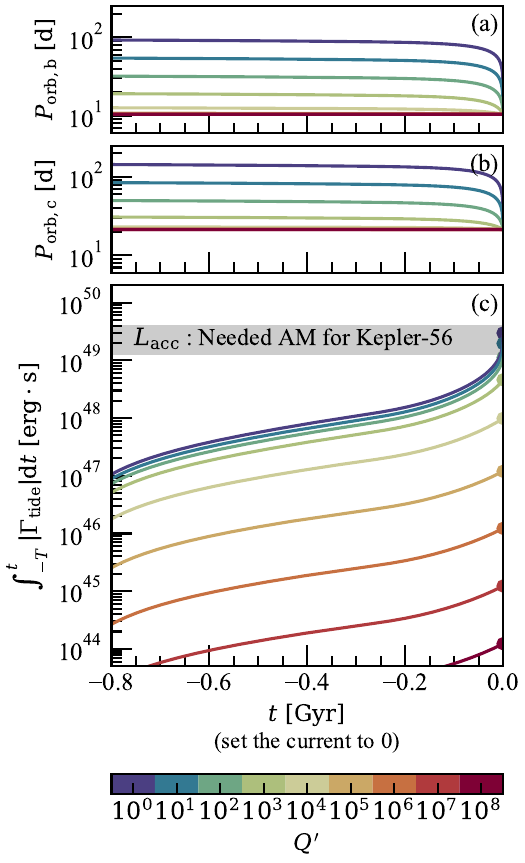}
 \caption{Time evolution of the orbital periods of Kepler-56 b and c ($P_\mathrm{orb, b}$ and $P_\mathrm{orb, c}$, respectively) and the amounts of maximised AM supply to Kepler-56 for the spin acceleration  ($L_\mathrm{calc, max}$) through the tidal interaction with Kepler-56 b and c at each $Q'$. The horizontal axis represents time, with the current time set to 0. The line colours correspond to the $Q'$ values (see the colour-bar). Panels (a) and (b) show the histories of $P_\mathrm{orb, b}$ and $P_\mathrm{orb, c}$ for each $Q'$, respectively. Panel (c) shows the histories of $L_\mathrm{calc, max}$ for each $Q'$. The values at $t=0$, which means the total amount of maximised AM supply, are highlighted with the circle symbols. For comparison, $L_\mathrm{acc}$ is shown as the grey shaded region (see Section \ref{subsec:meth-Lacc}). }
 \label{fig:evo-tides}
\end{figure}

\begin{figure}
 \includegraphics[width=\columnwidth]{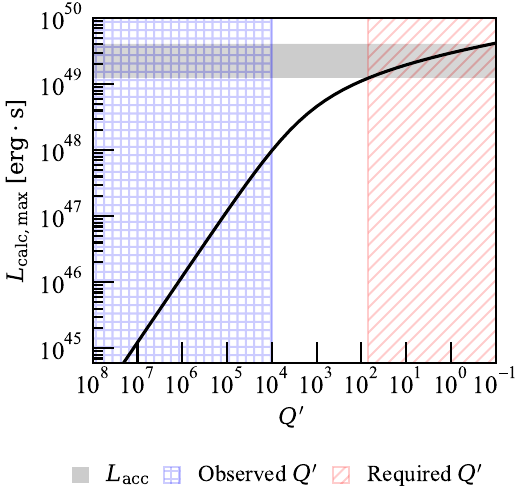}
 \caption{Relationship between the total amount of maximised AM supply and $Q'$. For comparison, $L_\mathrm{acc}$ is shown as the grey shaded region (see Section \ref{subsec:meth-Lacc}). The observational constraints on $Q'$ and the range of $Q'$ that can supply $L_\mathrm{acc}$ are indicated in blue and orange hatched regions, respectively.}
 \label{fig:res-tides}
\end{figure}

Fig.~\ref{fig:evo-tides} shows the time evolution of the orbital periods of Kepler-56 b and c and the amount of maximised AM supply for the spin acceleration through the tidal interaction with Kepler-56 b and c at each $Q'$. The coloured circles at the right end of panel (c) show the total amount of maximised AM supply ($L_\mathrm{calc, max}$; see Equation \ref{eq:def-Lcalcmax}) for each $Q'$. The relationship between $L_\mathrm{calc, max}$ and $Q'$ is summarized in Fig.~\ref{fig:res-tides}. 

By comparison with $L_\mathrm{acc}$ (grey shaded regions) in Fig.~\ref{fig:res-tides}, we can find that the extremely high tidal efficiency of $Q'=10^{-1}$-$10^2$ (red hatched region) is required to accelerate Kepler-56's spin. This is attributed to the required AM for spin-up is similar to or greater than the orbital AM of the current Kepler-56 b and Kepler-56 c. In such cases, it is necessary to place the initial orbit at a location far away from the current orbit, which is $P_\mathrm{orb} \gtrsim 50 \, \mathrm{d}$ in this case (see panels (a) and (b) in Fig.~\ref{fig:evo-tides}). Thus, the strong tidal efficiency is needed to bring the current position. Appendix \ref{app:Analytic} shows the explanation of this trend through the analytic formulation. 

The extremely strong tides with $Q' = 10^{-1}$–$10^{2}$ suggested here appear unreasonable, as they deviate from the observation-based range of $Q' = 10^{4}$–$10^{8}$ (blue hatched region in Fig.~\ref{fig:res-tides}). It should be noted, though, that such extremely strong tides in RGs cannot yet be completely ruled out, since the observational constraints are based on MS stars and subgiants. To constrain the tidal efficiency in RGs through future observations, transit-timing variations of RG systems are expected to be an effective approach. For Kepler-56, however, this is challenging because the system exhibits large perturbations arising from dynamical interactions \citep{Huber2013Sci}.

Nevertheless, the existence of Kepler-56 b/c and other hot Jupiters around RGs \citep{Temmink2023A&A, Pereira2024MNRAS} remains consistent with our results, as such systems would hardly be observable if extremely strong tides were universal. Theoretical calculations by \citet{Esseldeurs2024A&A} also support this view, suggesting that tides in the convective envelope do not reach such extreme efficiencies during the early RG stage for systems with stellar mass, evolutionary stage, and orbital periods similar to those of the Kepler-56 system.

The above considerations indicate that acceleration from a spin rate comparable to those of typical RGs to the rapid spin rate observed in Kepler-56 is difficult by tidal interactions between Kepler-56 b and Kepler-56 c alone. This suggests the necessity of a companion that originally existed inside the orbit of Kepler-56 b (see Sections \ref{subsec:res-case2} and \ref{subsec:companion}).

It should be emphasized that this conclusion based on the assumption of the slow spin rate of Kepler 56 at the initial conditions. This does not apply if Kepler-56 was spinning rapidly at the initial conditions. This case is discussed in Section \ref{subsec:rapid-spin}.

\subsection{Required conditions for engulfed hot Jupiter} \label{subsec:res-case2}

\begin{figure}
 \includegraphics[width=\columnwidth]{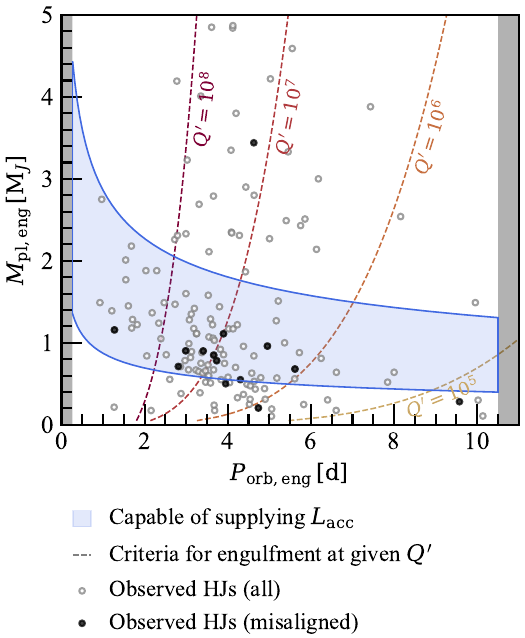}
 \caption{Requirements of $(M_\mathrm{pl,eng}, P_\mathrm{orb,eng})$ for engulfed HJ capable of reproducing the spin profile of Kepler-56 and comparison with observed HJs around $1.20-1.45 \, \mathrm{M}_\odot$ MS stars (circle symbols, see Section \ref{subsec:meth-scenario}). The blue shaded region corresponds to the requirement that $L_\mathrm{acc}$ can be supplied by its engulfment. The broken lines indicate the criteria for a planet to undergo engulfment at the given $Q'$. The area to the left of the broken line is the region where the planet can engulf at each $Q'$, and the colour of the broken line represents the value of $Q'$. HJs whose orbits are known to be misaligned are marked with filled circles, while those that are unknown are marked with open ones. The grey areas on the left and right sides indicate regions where HJ cannot exist due to the properties of the Kepler-56 system. The left side is the region where the orbit is inside the star in the MS stage, and the right side is the region where the orbit is outside the orbit of Kepler-56 b (see Section \ref{subsec:meth-scenario}).}
 \label{fig:res-engulf}
\end{figure}

The results of Section \ref{subsec:res-case1} simplify the interpretation of the requirement that engulfed HJ to reproduce Kepler-56, as discussed in Section \ref{subsec:meth-case2}. Fig.~\ref{fig:res-tides} shows that Kepler-56 b and c can only supply an AM of less than a few per cent of $L_\mathrm{acc}$ under the observation-based $Q'$ range, $Q'=10^4$-$10^8$. It means that the contributions of Kepler-56 b and c are subdominant under the assumption that an engulfed HJ supplies $L_\mathrm{acc}$.
Considering the AM supply of the engulfed HJ matches the initial orbital AM, the capability to supply $L_\mathrm{acc}$ for the engulfed HJ is determined mostly by the initial orbital AM and is almost independent of $Q'$. Thus, $Q'$ can be treated as the parameter that determines only whether the planet can engulf or not. We note that the calculation of Section \ref{subsec:meth-case2} confirms that the above considerations are correct. 

Therefore, for the engulfed HJ, we can consider separately the capability to supply $L_\mathrm{acc}$ and the requirement that the planet undergoes the engulfment during the post-MS stage. We can calculate the former by calculating the constraint that the initial orbit AM falls within the range of $L_\mathrm{acc}$. This result is not affected by the approximation of the expressions in Section \ref{subsec:meth-case1} or the uncertainty of $Q'$. On the other hand, the latter is discussed by the calculation in Section \ref{subsec:meth-case2}, which may be affected by the approximation performed in Section \ref{subsec:meth-case1} (specifically described in several paragraphs later). 

On the basis of the considerations above, Fig.~\ref{fig:res-engulf} shows the requirement of $(M_\mathrm{pl,eng}, P_\mathrm{orb,eng})$ for the engulfed HJ to reproduce Kepler-56. The blue shaded region in Fig.~\ref{fig:res-engulf} represents the region where the AM supply through the engulfment reaches $L_\mathrm{acc}$. The dashed line shows the criteria for the planetary engulfment at each $Q'$. We also plot the observed HJs around $1.20-1.45 \, \mathrm{M}_\odot$ MS stars (see Section \ref{subsec:meth-case2}). 

From Fig.~\ref{fig:res-engulf}, we can find that the region determined based on the capability of AM supply is consistent with the observed HJ population. It means that the engulfment of `typical' HJ could supply the AM necessary to reproduce Kepler-56. Specifically, an HJ with $M_\mathrm{pl,eng} \simeq 0.5$-$2 \, \mathrm{M}_\mathrm{J}$ and $P_\mathrm{orb,eng} \simeq 1$-$6 \, \mathrm{d}$ is preferred from the observation. The fact that the samples within that range include the systems with large spin-orbit obliquity (filled circle in Fig.~\ref{fig:res-engulf}) ensures that our scenario of an engulfment of misaligned HJ is possible. However, it should be noted that the actual required mass may be larger due to the effects ignored in this analysis (see Section \ref{subsec:addtional-effect}).

It can also be seen that engulfment of HJ in this range can be achieved with $Q' = 10^5-10^8$. This is consistent with the observational constraints of $Q'$. It should be noted that a stronger tidal efficiency is actually required since the calculations here are based on the assumption that tidal torque is always transferred from the planet to the star and works entirely on orbital decay (see Section \ref{subsec:meth-case2}). However, since an HJ in this region generally has orbital revolution faster than the stellar spin, the approximation should introduce only a minor error.

These results suggest that the engulfment of an HJ that originally existed inside Kepler-56 b can reproduce the spin structure of Kepler-56. In other words, Kepler-56 has the potential to serve as a benchmark for investigating the impact of planetary engulfment (see Sections \ref{subsec:chemical} and \ref{subsec:implication}).

\section{Discussion} \label{sec:discussion}

\subsection{Possibility of rapid initial-spin}
\label{subsec:rapid-spin}

The results of this study are based on the assumption that the rapid spin of Kepler-56 is due to AM supply in the post-MS stage (Sections \ref{subsec:meth-scenario} and \ref{subsec:meth-Lacc}). In this section, we consider the situation where spin-up is completed at the initial condition, which corresponds to $J_\mathrm{init} = J_\mathrm{obs}$ in Section \ref{subsec:meth-Lacc}. 

Specifically, we examine the efficiency of obliquity damping to reproduce Kepler-56 by Kepler-56 b and c. If we assume that the orbital periods of Kepler-56 b and c are constant and calculate Equations (\ref{eq:hi-evo-vector}), (\ref{eq:def-hi-norm}) and (\ref{eq:def-Gamma_tide-norm}), we find that $\simeq 10^{47} (10^4/Q') \, \mathrm{erg \cdot s}$ can be used for obliquity damping. It is emphasized that $Q'$ refers to the efficiency of obliquity damping, whereas the $Q'$ discussed in Sections \ref{sec:result} referred to the efficiency of tidal acceleration. To achieve an internal spin misalignment of 60 degrees or more in Kepler-56 (Section \ref{subsec:meth-scenario}), an AM supply of $J_\mathrm{damp}$, which satisfies $J_\mathrm{damp} \gtrsim J_\mathrm{obs} \gtrsim 10^{49} \, \mathrm{erg \cdot s}$ (Section \ref{subsec:meth-Lacc}), is required. Thus, the efficiency of obliquity damping corresponding to $Q' \lesssim 10^2$. \citet{Saunders2024AJ} show that $Q'$ for obliquity damping is roughly $10^4$ times smaller than that for tidal acceleration, that is to say, show that $Q' = 10^0 - 10^4$ for obliquity damping (cf. Section \ref{subsec:meth-case1}). Therefore, if Kepler-56 does not require any acceleration, obliquity damping by Kepler-56 b/c alone is sufficient to explain the spin profile of Kepler-56.

In such cases, however, we need to consider the process generating rapid spin. One scenario involves achieving rapid spin rate within the diversity of single stars. While magnetic braking reduces stellar AM, magnetic braking in F-type MS stars only occurs during the late phase of the MS stage \citep[][]{vanSadersPinsonneault2013ApJ, Beyer2024ApJ}. Therefore, it may be possible for stars to enter the RG stage while maintaining a rapid spin rate. However, this scenario raises several concerns: Kepler-56 deviates from the spin rate distribution of typical RGs (Fig.~\ref{fig:Prot_env_core}), which may represent the diversity at the end of the MS stage \footnote{This inference is based on the rough agreement on AM between few old F-type MS stars studied in asteroseismology \citep{Hall2021NatAs} and RG in Fig.~\ref{fig:Prot_env_core} \citep{Li2024A&A}. }; and Kepler-56's metal-rich composition deepens its surface convection zone, which enhances the efficiency of magnetic braking \citep{AmardMatt2020ApJ, See2024MNRAS}.

Another scenario involves AM transfer from a companion. As shown in Section \ref{subsec:companion}, the presence of Kepler-56 b/c makes binary interactions unlikely in the Kepler-56 system. Thus, the engulfment of a hot Jupiter again becomes a plausible scenario, in which case not only tidal forces but also mechanical scattering are candidates as mechanisms causing the the engulfment. As discussed in Section \ref{subsec:meth-scenario}, planetary engulfment during the MS stage is expected not to create internal spin misalignment via internal angular momentum transport. Thus, unlike the setting in this study, a scenario is suggested in which the stellar spin becomes rapid and globally misaligned after the engulfment of a hot Jupiter, with only the envelope later realigning during the post–MS phase. This implies that the engulfed hot Jupiter had an orbit inclined relative to the current orbits of Kepler-56 b/c. While this naively suggests a system hosting a single inclined hot Jupiter, it is unlikely to form or maintain such a configuration. A more plausible scenario involves the observed massive outer planet, Kepler-56 d: the orbits of the engulfed hot Jupiter and Kepler-56 b/c were initially aligned, but after the early engulfment, the orbits of Kepler-56 b/c gradually became inclined through dynamical interactions with Kepler-56 d before the star reached the post–MS stage \citep{Huber2013Sci,Li2014ApJ}. 

In any case, further investigation of F-type main sequence stars and hot Jupiter systems is requires for examining feasibility of rapid initial-spin.


\subsection{Scenario involving heavier companion}
\label{subsec:companion}

In this paper, we focused on the engulfment of HJ among merger/engulfment processes with a companion, obtaining the feasible scenario (Section \ref{subsec:res-case2}). The results are based on the assumption, which is considered reasonable in HJ, that all orbital AM is transferred to Kepler-56's spin AM through the engulfment. However, even through the merger process with a heavier companion such as a brown dwarf and low-mass star, Kepler-56 may be reproducible if the efficiency of AM supply in the merger process is not high. In fact, this situation is often considered in the common envelope process \citep[e.g.][]{Ivanova2013A&ARv}. 

While this efficiency is unknown, making it difficult to evaluate quantitatively, considerations about circumbinary planets provide severe constraints on the stellar-merger scenario. If Kepler-56 is the by-product of a stellar merger, Kepler-56 b and c would have originally been close-in circumbinary planets around a short-period binary. However, it is considered that such a system would be difficult to form from the perspectives of both formation process \citep{Moriwaki2004ApJ,Silsbee2015ApJ} and orbital stability \citep{Hamers2016MNRAS, Fleming2018ApJ}. In fact, no such system has ever been observed \citep{Armstrong2014MNRAS, Martin2015MNRAS}. That is to say, in order to reproduce the environment in which planets already exist around Kepler-56, it is preferable to assume that the original companion is lightweight, preferring an HJ. In this regard, the numerical simulation of long-term orbital evolution may provide quantitative constraints.

Additionally, the trend of chemical composition also disfavours the stellar-merger scenario. The carbon and nitrogen abundance of RGs, which is determined by the depth of the convective region and the rate of CNO burning, is strongly correlated with stellar mass and metallicity \citep[e.g.][]{Iben1964ApJ, Martig2016MNRAS, Bufanda2023ApJ}. If there is a significant mass supply, such as a stellar merger, these elements may deviate from the trends of analogous RGs. However, as shown in Section \ref{subsec:chemical}, Kepler-56 exhibits a consistent composition with analogous RGs.

\subsection{Impact of additional effect}
\label{subsec:addtional-effect}

This section discusses the impact on the results of effects neglected in this study, namely planetary atmospheric escape and magnetic braking.

\subsubsection{Atmospheric escape of hot Jupiter}
\label{subsubsec:escape}

In this study, the planetary mass was fixed as a constant (Section \ref{sec:method}). However, it is considered that the mass and radius of close-in planets decrease
through atmospheric escape by the stellar radiation \citep[e.g.][]{Owen2019AREPS}. This shifts the range of required planetary masses toward heavier masses than those shown in Fig.~\ref{fig:res-engulf}. 

Specifically, the criterion for whether a planet can maintain its envelope is thought to be whether it is heavier than $\sim 1 \, \mathrm{M}_\mathrm{J}$ \citep{KurokawaNakamoto2014ApJ}. Based on this criterion, under an HJ which satisfies the conditions described in Section \ref{subsec:res-case2}, that with $\gtrsim 1 \, \mathrm{M}_\mathrm{J}$ may not be significantly affected and that with $\lesssim 1 \, \mathrm{M}_\mathrm{J}$ may be unable to achieve a sufficient AM supply. As a result, the engulfment of the more massive and closer HJ is considered to be preferable. More detailed investigations require calculations that simultaneously solve the evolution of planetary atmosphere for each individual initial condition. 

\subsubsection{Magnetic Braking}
\label{subsubsec:mag-braking}


In this study, AM loss due to the magnetic braking was neglected in Section \ref{subsec:meth-case1}. However, the magnetic braking may need to be considered for rapidly-spinning red giants like Kepler-56 \citep{Privitera2016bA&A, Ong2024ApJ}. If the discussion for main-sequence stars is extended to red giants, the efficiency of magnetic braking is expected to be governed by the Rossby number, defined as the ratio of the rotation period to the convective turnover time; Magnetic braking operates when the Rossby number is below unity \citep{vanSaders2016Natur, Tokuno2023MNRAS, Saunders2024ApJ, Metcalfe2025ApJ}, and its efficiency increases as the Rossby number decreases \citep[e.g.][]{vanSadersPinsonneault2013ApJ, Matt2015ApJ}. Empirical estimate of \citet[][]{Lehtinen2020NatAs} suggests that the convective turnover time of Kepler-56 is $\sim 1 \, \mathrm{yr}$, corresponding to a Rossby number of $\sim 0.2$. Therefore, the magnetic braking likely works on Kepler-56 efficiently.

If the AM loss due to magnetic braking is effective, Kepler-56 would have been spinning more rapid than the current spin rate. Thus, the required angular momentum to accelerate Kepler-56 ($L_\mathrm{acc}$ in Section \ref{subsec:meth-Lacc}) increases. This increases the required tidal efficiency for acceleration by Kepler-56 b/c in Section \ref{subsec:res-case1} and increases the required mass for engulfed hot Jupiters in Section \ref{subsec:res-case2}. However, it is difficult to determine the extent of this increase, as it depends on the efficiency of magnetic braking, which is not yet well established in the RG stage \citep[e.g.][]{Ceillier2017A&A}, and on the timing of absorption, which is difficult to investigate.

\subsection{Chemical abundance of Kepler-56}
\label{subsec:chemical}

\begin{figure*}
 \includegraphics[width=2\columnwidth]{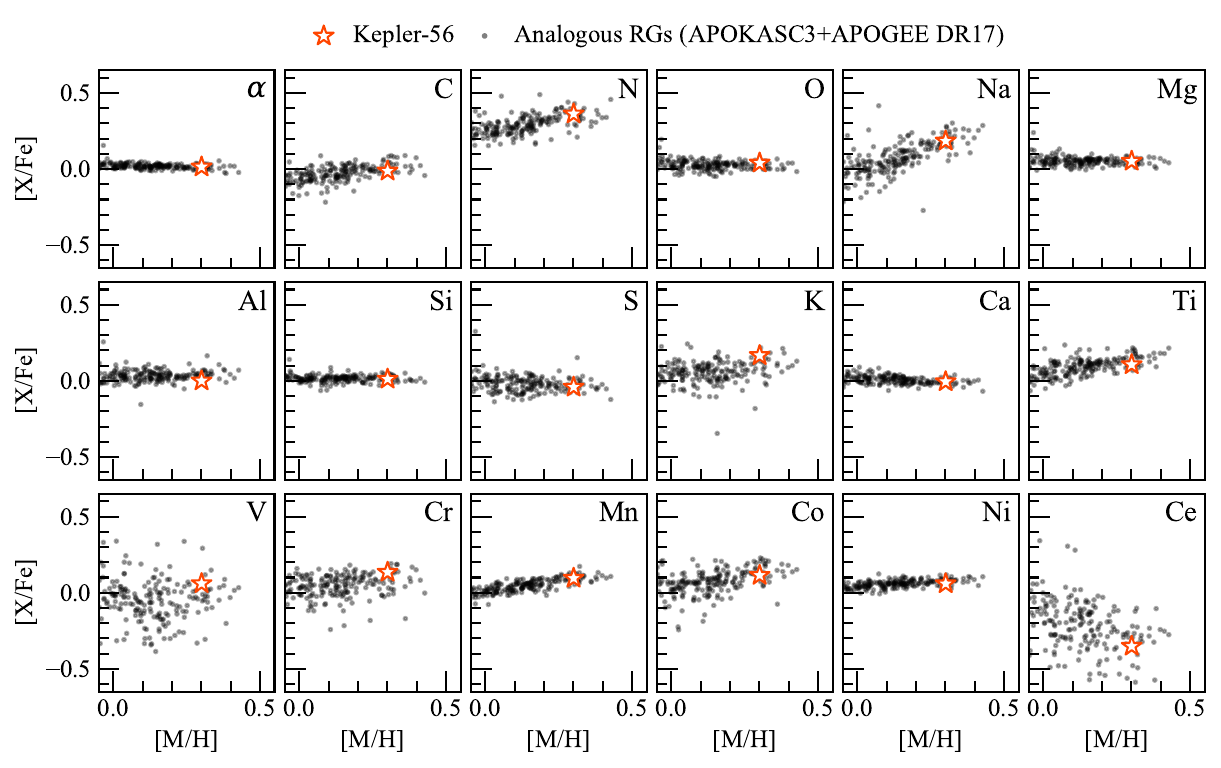}
 \caption{Relationship between metallicity [M/H] and abundances [X/Fe] for each element (see the upper-right corner of each panel). For each panel, the values of Kepler-56 and analogues RGs (1.20-1.45 $ \mathrm{M}_\odot$ for mass and 3.0-6.0 $\mathrm{R}_\odot$ for radius) are represented by the black circles and the orange star-shaped symbol, respectively. All values are retrieved in the catalogues of APOGEE DR17 \citep{Abdurro'uf2022ApJS} and APOKASC3 \citep{Pinsonneault2025ApJS}. }
 \label{fig:Abundance}
\end{figure*}

It is suggested that planetary engulfment may alter the chemical abundance at the stellar surface \citep{Zuckerman2010ApJ, Oh2018ApJ, Spina2021NatAs, Liu2024Natur}. Here, we examine the chemical characteristics of Kepler-56 relative to the analogues RGs. From the APOKASC3 catalogue \citep{Pinsonneault2025ApJS},
we extracted RGs with $1.20-1.45 \, \mathrm{M}_\odot$ for mass and $R_\mathrm{s}=3.0-6.0 \, \mathrm{R}_\odot$ for radius. After that, we retrieved the chemical abundance for each star from the results of APOGEE DR17 \citep{Abdurro'uf2022ApJS}.

Fig.~\ref{fig:Abundance} shows the relationship between the metallicity and the abundance for each element. We find that the surface composition of Kepler-56 falls within the range of dispersion of analogues RGs. This result is consistent with the claim that compositional changes due to the planetary engulfment are comparable to or smaller than primordial dispersion \citep{Behmard2023MNRASa}. 

\citet{Tayar2022ApJ} performed a similar analysis on another rapidly rotating RG, KIC 9267654, and also showed that no peculiar chemical composition was detected. If Kepler-56 or KIC 9267654 are actually experiencing the planetary engulfment, this trend suggests that it may be difficult to detect signatures of planetary engulfment in their chemical compositions. Thus, in order to investigate the trace of engulfment in RGs from their composition, it may be necessary to use a method that is less affected by primordial dispersion, e.g. the relative differences with a binary companion \citep{Behmard2023MNRASb}.

\subsection{Implication for other systems}
\label{subsec:implication}

This section discusses the implications of our findings for the formation process of other RGs with rapid spin or internal spin misalignment.

Many rapidly spinning RGs have been discovered by photometric or spectroscopic analyses \citep{Carlberg2011ApJ, Carlberg2013AN, Tayar2015ApJ, Tayar2022ApJ, Ceillier2017A&A,Patton2024MNRAS, Dhanpal2025arXiv}. Engulfment of HJ is considered important as a possible formation process for such objects, but remains suggestive because of  indistinguishability from other scenarios. In this respect, Kepler-56 is a valuable object that is strongly suggested to have undergone the engulfment of HJ. This is attributed to further peculiarities of Kepler-56: its internal spin misalignment that is difficult to form in a single star, and the presence of a nearby planet that imposes strong constraints (Sections \ref{subsec:meth-scenario} and \ref{subsec:companion}). This suggestion that one of the rapidly spinning RGs may be formed by planetary engulfment highlights the importance of planetary engulfment in the AM evolution of RGs. 

On the other hand, it is unlikely that all rapidly spinning RGs were created by planetary engulfment. The proportion of rapidly spinning RGs is several per cent of all RGs \citep{Carlberg2011ApJ, Tayar2015ApJ, Patton2024MNRAS}, while less than 0.5 per cent of stars have HJs \citep{Beleznay2022MNRAS}. Furthermore, some fast-rotating RGs have such enormous AM that they cannot be supplied by the planetary engulfment ($v \sin i > 20 \, \mathrm{km/s}$). Therefore, most rapidly rotating RGs should be considered to be by-products of binary interaction \citep{Tayar2015ApJ, Patton2024MNRAS}.

Currently, the only star with an internal spin axis misalignment is Kepler-56. However, such stars are considered to be formed at a certain proportion through the interaction with a companion with a misaligned orbit. In fact, a misaligned HJ is common around A-type or F-type MS stars, which may become systems like Kepler-56 in the future \citep[][see also Appendix \ref{app:Obliquity}]{Winn2010ApJ, Albrecht2011ApJ}. Due to the difficulty of maintaining the misalignment with the heavier companion, the stars resulting from planetary engulfment, as with Kepler-56, may be observed preferentially. Thus, in order to understand the effects of planetary engulfment, it is important to investigate objects with internal spin misalignment, which may be achieved  through future analyses, such as the PLAnetary Transits and Oscillations of stars (PLATO) mission \citep{Rauer2014ExA, Rauer2025ExA}.

\section{Summary}
\label{sec:conclusion}

Kepler-56 has provoked discussion due to its unique spin structure, namely, rapid envelope spin and internal spin misalignment, revealed by recent asteroseismic analyses \citep{Ong2025ApJ}. These features strongly suggest past interactions with a companion star, making this an ideal object for verifying the stellar angular momentum (AM) evolution through such interactions. This paper aims to investigate the feasible scenario to reproduce the spin structure of Kepler-56 through the AM transfers with planets. In particular, we examined the following two scenarios from the perspective of whether acceleration from the spin rate observed in typical RGs to Kepler-56' spin rate could be reproduced: (1) AM supply via tides with the observed planets, Kepler-56 b and c, and (2) AM supply through the engulfment of a hot Jupiter (HJ) that originally existed inside the observed planets. By solving simplified equations for AM evolution, we constrained the minimum tidal efficiency for acceleration through tidal interaction with the observed planets and derived the properties of HJ for acceleration through its engulfment.

The main results of this study are as follows. First, we showed that acceleration of Kepler-56 solely through tidal interaction with the known close-in planets requires a tidal efficiency far exceeding observational constraints. This result suggests a scenario in which the engulfment of a companion that originally existed supplies AM under the assumption that Kepler 56's spin rate was originally equivalent to those of typical RGs. Second, we demonstrated that an HJ with a mass of 0.5–2 Jupiter masses and a spin period of approximately 1–6 days, if present during the main-sequence stage, can accelerate Kepler-56 without conflicting with the observations. It should be emphasized that, if Kepler-56 was initially rapidly spinning, it might be possible to reproduce Kepler-56 solely through the obliquity damping of Kepler-56 b/c. However, in that case, some scenario would be necessary to achieve the initial rapid spin rate, including planetary engulfment.

These results highlight the importance of Kepler-56 as the benchmark star showing traces of planetary engulfment and demonstrate the significance of the planetary engulfment for the stellar AM evolution.

\section*{Acknowledgements}
I thank the referee for constructive comments that led to an improvement of the quality of our study. I also thank Takeru K. Suzuki, Ryo Tazaki, Masao Takata, James Fuller, Akihiko Fukui and Yugo Kawai for insightful comments that improved the presentation of this work. This work is supported by Grants-in-Aid for Scientific Research from the MEXT/JSPS of Japan (KAKENHI grant No. JP24KJ0605). 


\section*{Data availability}
The data underlying this article will be shared on reasonable request to the corresponding author. 



\bibliographystyle{mnras}
\bibliography{example} 



\appendix



\section{MESA Calculation}
\label{app:MESA}

\begin{table}
	\centering
	\caption{The setting of the stellar evolution code \textsc{mesa} in this paper.
	}
	\label{tab:mesa_setting}
	\begin{tabular}{ll} 
		\hline
		Control name & Our setting \\ 
		\hline
		\texttt{initial\_mass} & $1.19$, $1.26$, $1.32$, $1.38$ or $1.45 \, \mathrm{M}_{\odot}$ \\
		\texttt{initial\_Z} & $Z=0.027 ^{*a}$ \\ 
		\texttt{use\_Type2\_opacities} & True  \\
        \texttt{kap\_file\_prefix} & 'gs98' \\
	\texttt{default\_net\_name} & 'pp\_and\_cno\_extras'   \\
		\texttt{atm\_T\_tau\_relation} & 'Eddington' \\
		\texttt{mixing\_length\_alpha} & $\alpha_\mathrm{MLT}=2.0$ \\
        \texttt{MLT\_option} & 'TDC' \\
		\texttt{do\_element\_diffusion} & True \\
        \texttt{use\_Ledoux\_criterion} & False 
		\\
		\hline
        \multicolumn{2}{l}{$^{*a}$ It corresponds to [Fe/H]=0.20 at $Z_\odot=0.172$ \citep{Hidargo2018ApJ}}
	\end{tabular}
\end{table}

In this paper, we calculate the evolutionary tracks of Kepler-56 by the stellar evolution code \textsc{mesa} (r24.08.1). Our settings of major parameters are listed in Table \ref{tab:mesa_setting}. We do not include an overshooting process. ignore the mass-loss by stellar winds because its contribution is negligible. As a supplement, calculations were also performed for [Fe/H] = 0.05 and 0.35 (see Table \ref{tab:values_K56} but there are little impacts on our results. 

\section{List of misaligned hot jupiters}
\label{app:Obliquity}

Table \ref{tab:obliquity} shows a list of HJs around F-type MS stars with a projected spin-orbit obliquity of more than 50 degrees (Section \ref{subsec:meth-case2} and Fig.~\ref{fig:res-engulf}). These systems are expected to create RGs with spin structures like Kepler-56 in the future. We note that none of these systems were found to be multiple planetary systems like the Kepler-56 system.

\begin{table}
    \centering
    \caption{List of HJs around F-type MS stars with spin-orbit misalignment.}
    \label{tab:obliquity}
    \begin{tabular}{lcccc} 
    \hline
    Planet Name & $P_\mathrm{orb} \, [\mathrm{d}]$ & $M_\mathrm{pl} \, [\mathrm{M}_\odot]$  & $\lambda \, [\mathrm{deg}]$ & Source$^{*a}$  \\
    \hline
    WASP-121 b & 1.275 & 1.16 & $87^{+0.4}_{-0.4}$ & (1) \\
    HAT-P-30 b & 2.811 & 0.71 & $74^{+9}_{-9}$ & (2) \\
    KELT-4 A b & 2.990 & 0.90 & $80^{+25}_{-22}$ & (3) \\
    WASP-180 A b & 3.409 & 0.90 & $-157^{+2}_{-2}$ & (4) \\
    WASP-79 b & 3.662 & 0.85 & $-95^{+1}_{-1}$ & (5) \\
    WASP-17 b & 3.735 & 0.78 & $-149^{+5}_{-4}$ & (6) \\
    CoRoT-19 b & 3.897 & 1.11 & $-52^{+27}_{-22}$ & (7) \\
    WASP-94 A b & 3.950 & 0.45 & $151^{+16}_{-23}$ & (8) \\
    WASP-60 b & 4.305 & 0.55 & $-129^{+17}_{-17}$ & (9) \\
    HAT-P-14 b & 4.628 & 3.44 & $189^{+5}_{-5}$ & (10) \\
    KELT-11 b & 4.736 & 0.21 & $-78^{+2}_{-2}$ & (11) \\
    WASP-7 b & 4.955 & 0.96 & $86^{+6}_{-6}$ & (12) \\
    CoRoT-36 b & 5.617 & 0.68 & $276^{+11}_{-11}$ & (13) \\
    TOI-1842 b & 9.574 & 0.28 & $-68^{+21}_{-15}$ & (14) \\
    \hline
    \multicolumn{5}{l}{$^{*a}$ (1)  \citet{Bourrier2020A&A}; (2) \citet{Johnson2011ApJ};  } \\
    \multicolumn{5}{l}{(3) \citet{Knudstrup2024A&A}; (4) \citet{Temple2019MNRAS};  } \\
    \multicolumn{5}{l}{(5) \citet{Brown2017MNRAS}; (6) \citet{Triaud2010A&A}; } \\
    \multicolumn{5}{l}{(7) \citet{Guenther2012A&A}; (8) \citet{Neveu-VanMalle2014A&A}; } \\
    \multicolumn{5}{l}{(9) \citet{Mancini2018A&A} ; (10) \citet{Winn2011AJ}; } \\
    \multicolumn{5}{l}{(11) \citet{Mounzer2022A&A}; (12) \citet{Albrecht2012ApJ}; } \\
    \multicolumn{5}{l}{(13) \citet{Sebastian2022MNRAS}; (14) \citet{Hixenbaugh2023ApJ}; } \\
    \end{tabular} 
\end{table}

\section{Integration of $\Gamma_\mathrm{tide}$}
\label{app:Analytic}

In this study, we conduct numerical integration of Equation (\ref{eq:def-Lcalcmax}) based on the time evolution of $R_\mathrm{s}$ (Section \ref{subsec:meth-case1}). However, assuming $R_\mathrm{s}$ to be constant, the integration of Equation (\ref{eq:def-Lcalcmax}) can be solved analytically. This integral helps in understanding the behaviour shown in Fig.~\ref{fig:res-tides}.

Considering the initial condition $n_i(0)=n_{i,0}$ at $t=0$, we obtain
\begin{align}
    n_i(t) &= n_{i,0} \left( 1 - 13 \frac{|\mathbf{\Gamma}_{\mathrm{tide},i}|_{t=0}}{|\bm{h}_i|_{t=0}} t \right)^{-3/13} \notag \\
    & = n_{i,0} \left( 1 - \frac{Q'_{\mathrm{ref},i}}{Q'} \frac{t}{T} \right)^{-3/13}, \label{eq:n-analytic}
\end{align}
where $T$ is an arbitrarily selectable time interval and $Q'_\mathrm{ref}$ is the reference tidal quality factor defined as
\begin{align}
    Q'_\mathrm{ref} &= \frac{117 R_\mathrm{s}^5 M_\mathrm{pl}  n_0^{13/3} T}{ 4 G^{5/3} M_\mathrm{s}(M_s+M_\mathrm{pl})^{5/3}} \notag \\
    & \simeq 2.2 \times 10^3 \left( \frac{n_0/2\pi}{0.1 \, \mathrm{d}^{-1}} \right)^{13/3} \left( \frac{M_\mathrm{s}}{1 \, \mathrm{M}_\odot} \right)^{-8/3} \left( \frac{R_\mathrm{s}}{1 \, \mathrm{M}_\odot} \right)^{5} \times \notag \\
    & \qquad  \left( \frac{M_\mathrm{pl,i}}{1 \, \mathrm{M_J}} \right) \left( \frac{T}{1 \, \mathrm{Gyr}} \right) \quad (\text{at} \, M_\mathrm{s} \gg M_\mathrm{pl,i}). \label{eq:def-Qref}
\end{align}
From Equations (\ref{eq:hi-evo-vector}), (\ref{eq:def-hi-norm}), (\ref{eq:def-Lcalcmax}), (\ref{eq:n-analytic}) and (\ref{eq:def-Qref}), the amount of AM supply during $t=[0,T] $ is given by
\begin{gather}
    L_\mathrm{calc,max} = \sum_i |\bm{h}_i|_{t=0} \left[ \left( 1 + \frac{Q'_{\mathrm{ref},i}}{Q'} \right)^{1/13} - 1 \right]. \label{eq:gamma_analytic}
\end{gather}
Equation (\ref{eq:gamma_analytic}) closely approximates the relationship in Fig.~\ref{fig:res-tides} when the values of Kepler-56, $R_\mathrm{s} \simeq 3 \, \mathrm{R}_\odot$ and $T \simeq 1 \, \mathrm{Gyr}$, are substituted. We note that the values of $Q'_\mathrm{ref}$  for Kepler-56 b and c in that case are roughly $7 \times 10^3$ and $2 \times 10^4$, respectively.

It can be seen that AM supply is proportional to $Q'$ when $Q'\gg Q'_\mathrm{ref, i}$, and that its dependence on $Q'$ is very weak when $Q' \lesssim Q'_\mathrm{ref, i}$, which is shown in Fig.~\ref{fig:res-tides}. It indicates the difficulty of supplying AM exceeding the current orbital AM, $|\bm{h}_i|_{t=0}$ (see Section \ref{subsec:res-case1}).


\bsp	
\label{lastpage}
\end{document}